\documentclass[12pt]{article}
\usepackage{epsf}
\makeindex

\def\keywords#1{\small\centerline{\bf Key Words}\vspace{5mm}\centerline{\parbox{14cm}{#1}}}

\def\gtapprox{\buildrel{\lower.7ex\hbox{$>$}}\over
                       {\lower.7ex\hbox{$\sim$}}}
\def\nq{\hspace{-1em}}

\def\ignore#1{}

\def\odt{{\textstyle{1\over 2}}}

\def\hbar{h\!\!\!\!^{-}\,}

\def\eps{\varepsilon}
\def\beq{\begin{equation}}
\def\eeq{\end{equation}}
\def\beqn{\begin{displaymath}}
\def\eeqn{\end{displaymath}}
\def\bqa{\begin{equation}\begin{array}{c}}
\def\eqa{\end{array}\end{equation}}
\def\bqan{\begin{displaymath}\begin{array}{c}}
\def\eqan{\end{array}\end{displaymath}}

\def\length{{l}}

\begin{document}


\begin{titlepage}


\begin{center}
    {\small Technical Report IDSIA-16-00, 3 March 2001 \\
            {ftp://ftp.idsia.ch/pub/techrep/IDSIA-16-00.ps.gz}
    }\\[2cm]
  {\LARGE\sc\sc\hrule height1pt \vskip 2mm
          The Fastest and Shortest Algorithm \\
           for All Well-Defined Problems\footnotemark
   \vskip 5mm \hrule height1pt} \vspace{1.5cm}
  {\bf Marcus Hutter}                    \\[1cm]
  {\rm IDSIA, Galleria 2, CH-6928 Manno-Lugano, Switzerland}  \\
  {\rm\footnotesize marcus@idsia.ch \qquad
      http://www.idsia.ch/$^{_{_\sim}}\!$marcus} \\[1cm]
\end{center}
\footnotetext{Published in the {\it
International Journal of Foundations of Computer Science},
Vol. 13, No. 3, (June 2002) 431--443. Extended version of
{\it An effective Procedure for Speeding up Algorithms} (cs.CC/0102018)
presented at Workshops MaBiC-2001 \& TAI-2001.}


\begin{abstract}
An algorithm $M$ is described that solves any well-defined problem
$p$ as quickly as the fastest algorithm computing a solution to
$p$, save for a factor of 5 and low-order additive terms. $M$
optimally distributes resources between the execution of provably
correct $p$-solving programs and an enumeration of all proofs,
including relevant proofs of program correctness and of time
bounds on program runtimes. $M$ avoids Blum's speed-up theorem by
ignoring programs without correctness proof. $M$ has broader
applicability and can be faster than Levin's universal search, the
fastest method for inverting functions save for a large
multiplicative constant. An extension of Kolmogorov complexity and
two novel natural measures of function complexity are used to show
that the most efficient program computing some function $f$ is
also among the shortest programs provably computing $f$.
\end{abstract}

\end{titlepage}

\section{Introduction \& Main Result}\label{secInt}
Searching for fast algorithms to solve certain problems is a
central and difficult task in computer science. Positive results
usually come from explicit constructions of efficient algorithms
for specific problem classes. A wide class of problems can be
phrased in the following way. Given a formal specification of a
problem depending on some parameter $x\!\in\!X$, we are interested
in a fast algorithm computing solution $y\!\in\!Y$. This means
that we are interested in a fast algorithm computing
$f\!:\!X\!\to\!Y$, where $f$ is a formal (logical, mathematical,
not necessarily algorithmic), specification of the problem.
Ideally, we would like to have the fastest algorithm, maybe apart
from some small constant factor in computation time.
Unfortunately, Blum's Speed-up Theorem \cite{Blum:67,Blum:71}
shows that there are problems for which an (incomputable) sequence
of speed-improving algorithms (of increasing size) exists, but no
fastest algorithm.

In the approach presented here, we consider only those algorithms
which {\it provably} solve a given problem, and have a fast (i.e.\
quickly computable) time bound. Neither the programs themselves,
nor the proofs need to be known in advance. Under these
constraints we construct the asymptotically fastest algorithm
save a factor of 5 that solves any well-defined problem $f$.

\hfill

\begin{samepage}
{\sc Theorem 1.}
{\sl Let $p^*$ be a given algorithm computing $p^*(x)$ from x, or,
more generally, a specification of a function. Let $p$ be any
algorithm, computing provably the same function as $p^*$ with
computation time provably bounded by the function
$t_p(x)$ for all $x$. $time_{t_p}(x)$ is the
time needed to compute the time bound $t_p(x)$.
Then the algorithm $M_{p^*}$ constructed in Section
\ref{secFast} computes $p^*(x)$ in time
\beqn
  time_{M_{p^*}}(x) \;\leq\;
  5\!\cdot\!t_p(x) +
  d_p\!\cdot\!time_{t_p}(x) +
  c_p
\eeqn
with constants $c_p$ and $d_p$ depending on $p$ but not on
$x$. Neither $p$, $t_p$, nor the proofs need to be
known in advance for the construction of $M_{p^*}(x)$.
} 
\end{samepage}

\hfill

Known time bounds for practical problems can often be computed
quickly, i.e. $time_{t_p}(x)/time_p(x)$ often converges very
quickly to zero. Furthermore, from a practical point of view, the
provability restrictions are often rather weak. Hence, we have
constructed for every problem a solution, which is asymptotically
only a factor $5$ slower than the (provably) fastest algorithm!
There is no large multiplicative factor and the problems are not
restricted to inversion problems, as in Levin's algorithm (see
section \ref{secLevin}). What somewhat spoils the practical
applicability of $M_{p^*}$ is the large additive constant $c_p$,
which will be estimated in Section \ref{secTime}.

An interesting and counter-intuitive consequence of Theorem 1,
derived in Section \ref{secAIT}, is that the fastest program that
computes a certain function is also among the shortest programs
that provably computes this function. Looking for larger programs
saves at most a finite number of computation steps, but cannot
improve the time order.

In section \ref{secLevin} we review Levin search and the universal
search algorithms {\sc simple} and {\sc search}, described in
\cite{Li:97}. We point out that {\sc simple} has the same asymptotic
time complexity as {\sc search} not only w.r.t.\ the problem
instance, but also w.r.t.\ to the problem class.
In Section \ref{secDiscuss} we
elucidate Theorem 1 and the range of applicability on
an example problem unsolvable
by Levin search. In Section \ref{secFast} we give formal
definitions of the expressions {\it time}, {\it proof}, {\it
compute}, etc., which occur in Theorem 1, and define the fast
algorithm $M_{p^*}$. In Section
\ref{secTime} we analyze the algorithm $M_{p^*}$, especially its
computation time, prove Theorem 1, and give upper bounds for the
constants $c_p$ and $d_p$. Subtleties regarding the underlying
machine model are briefly discussed in Section \ref{secAssume}. In
Section \ref{secAIT} we show that the fastest program computing a
certain function is also among the shortest programs provably
computing this function. For this purpose, we extend the
definition of the Kolmogorov complexity of a string and define two
new natural measures for the complexity of functions and programs.
Section \ref{secGeneral} outlines generalizations of Theorem 1 to
i/o streams and other time-measures. Conclusions are given in
Section \ref{secConc}.

\section{Levin Search}\label{secLevin}
Levin search is one of the few rather general speed-up algorithms.
Within a (typically large) factor, it is the fastest algorithm for
inverting a function $g\!:\!Y\to\!X$, if $g$ can be evaluated quickly
\cite{Levin:73,Levin:84}. Given $x$, an inversion algorithm $p$
tries to find a $y\!\in\!Y$, called g-witness for $x$, with
$g(y)\!=\!x$. Levin search just runs and verifies the result of
{\em all} algorithms $p$ in parallel with relative computation
time $2^{-\length(p)}$; i.e.\ a time fraction $2^{-\length(p)}$ is
devoted to execute $p$, where $l(p)$ is the length of program $p$
(coded in binary). Verification is necessary since the output of
{\em any} program can be {\em anything}. This is the reason why
Levin search is only effective if a fast implementation of $g$ is
available. Levin search halts if the first $g$-witness has been
produced and verified. The total computation time to find a
solution (if one exists) is bounded by
$2^{\length(p)}\!\cdot\!time^+_p(x)$. $time^+_p(x)$ is the runtime
of $p(x)$ {\em plus} the time to verify the correctness of the
result ($g(p(x))\!=\!x$) by a {\em known} implementation for $g$.

Li and Vit\'anyi \cite[p503]{Li:97} propose a very simple variant,
called {\sc simple}, which runs all programs $p_1p_2p_3\ldots$ one
step at a time according to the following scheme: $p_1$ is run
every second step, $p_2$ every second step in the remaining unused
steps, $p_3$ every second step in the remaining unused steps, and
so forth, i.e.\ according to the sequence of indices
$121312141213121512\ldots$. If $p_k$ inverts $g$ on $x$ in
$time_{p_k}(x)$ steps, then {\sc simple} will do the same in {\em
at most} $2^k time^+_{p_k}(x)+2^{k-1}$ steps. In order to improve
the factor $2^k$, they define the algorithm {\sc search}, which
runs all $p$ (of length less than $i$) for $2^i2^{-l(p)}$ steps in
phase $i=1,2,3,\ldots$, until it has inverted $g$ on $x$. The
computation time of {\sc search} is bounded by
$2^{K(k)+O(1)}time^+_{p_k}(x)$, where
$K(k)\!\leq\!l(p_k)\!\leq\!2\log k$ is the Kolmogorov complexity
of $k$. They suggest that {\sc simple} has worse asymptotic
behaviour w.r.t.\ $k$ than {\sc search}, but actually this is not
the case.

In fact, {\sc simple} and {\sc search} have the same asymptotics
also in $k$,
because {\sc search} itself is an algorithm with some index
$k_{\mbox{\tiny SEARCH}}=\!O(1)$.
Hence, {\sc simple} executes {\sc search} every
$2^{k_{\mbox{\tiny SEARCH}}}$-th step, and can at most be a
constant (in $k$ and $x$) factor $2^{k_{\mbox{\tiny
SEARCH}}}=O(1)$ slower than {\sc search}. However, in practice,
{\sc search} should be favored, because also constants matter, and
$2^{k_{\mbox{\tiny SEARCH}}}$ is rather large.

Levin search can be modified to handle time-limited optimization
problems as well \cite{Solomonoff:86}. Many, but not all
problems, are of inversion or optimization type. The matrix
multiplication example (section \ref{secDiscuss}), the {\em
decision} problem SAT \cite[p503]{Li:97}, and reinforcement
learning \cite{Hutter:00f}, for instance, cannot be brought into
this form. Furthermore, the large factor $2^{\length(p)}$ somewhat
limits the applicability of Levin search. See
\cite[pp518-519]{Li:97} for a historical review and further
references.

Levin search in program space cannot be used directly in $M_{p^*}$
for computing $p^*$ because we have to decide somehow whether a
certain program solves our problem or computes something else. For
this, we have to search through the space of proofs. In order to
avoid the large time-factor $2^{l(p)}$, we also have to search
through the space of time-bounds. Only {\em one} (fast) program
should be executed for a significant time interval. The algorithm
$M_{p^*}$ essentially consists of 3 interwoven algorithms: {\em
sequential} program execution, sequential search through proof
space, and Levin search through time-bound space. A tricky
scheduling prevents performance degradation from computing slow
$p$'s before {\it the} $p$ has been found.

\section{Applicability of the Fast Algorithm $M_{p^*}$}\label{secDiscuss}
To illustrate Theorem 1, we consider the problem of multiplying
two $n\times n$ matrices. If $p^*$ is the standard algorithm for
multiplying two matrices\footnote{Instead of interpreting $R$ as
the set of real numbers one might take the field
$I\!\!F_2=\{0,1\}$ to avoid subtleties arising from large numbers.
Arithmetic operations are assumed to need one unit of time.}
$x\!\in\!R^{n\cdot n}\!\times\!R^{n\cdot n}$ of size
$\length(x)\!\sim\!n^2$, then $t_{p^*}(x)\!:=\!2n^3$ upper bounds
the true computation time $time_{p^*}(x)\!=\!n^2(2n-1)$. We know
there exists an algorithm $p'$ for matrix multiplication with
$time_{p'}(x)\leq t_{p'}(x)\!:=c\!\cdot\!n^{2.81}$
\cite{Strassen:69}. The time-bound function (cast to an integer)
can, as in many cases, be computed very quickly,
$time_{t_{p'}}(x)=O(log^2 n)$. Hence, using Theorem 1, also
$M_{p^*}$ is fast, $time_{M_{p^*}}(x)\leq 5c\!\cdot\!n^{2.81}+O(log^2
n)$. Of course, $M_{p^*}$ would be of no real use if $p'$ is
already the fastest program, since $p'$ is known and could be used
directly. We do not know however, whether there is an algorithm
$p''$ with $time_{p''}(x)\leq d\!\cdot\!n^2log\,n$, for instance.
But if it does exist, $time_{M_{p^*}}(x)\leq
5d\!\cdot\!n^2log\,n\!+\!O(1)$ for all $x$ is guaranteed.

The matrix multiplication example has been chosen for specific
reasons. First, it is not an inversion or optimization problem
suitable for Levin search. The computation time of Levin search is
lower-bounded by the time to verify the solution (which is at
least $c\!\cdot\!n^{2.81}$ to our knowledge) multiplied with the
(large) number of necessary verifications. Second, although matrix
multiplication is a very important and time-consuming issue, $p'$
is not used in practice, since $c$ is so large that for all
practically occurring $n$, the cubic algorithm is faster. The same
is true for $c_p$ and $d_p$, but we must admit that although $c$
is large, the bounds we obtain for $c_p$ and $d_p$ are tremendous.
On the other hand, even Levin search, which has a tremendous
multiplicative factor, can be successfully applied
\cite{Schmidhuber:97nn,Schmidhuber:97bias}, when handled with
care. The same should hold for Theorem 1, as will be discussed.
We avoid the $O()$ notation as far as possible, as it can be
severely misleading (e.g.\ $10^{42}=O(1)^{O(1)}=O(1)$). This work
could be viewed as another $O()$ warning showing, how important
factors, and even subdominant additive terms, are.

An obvious time bound for $p$ is the actual computation time
itself. An obvious algorithm to compute $time_p(x)$ is to count
the number of steps needed for computing $p(x)$. Hence, inserting
$t_p\!=\!time_p$ into Theorem 1 and using
$time_{time_p}(x)\!\leq\!time_p(x)$, we see that the computation
time of $M_{p^*}$ is optimal within a multiplicative constant
$(d_p+5)$ and an additive constant $c_p$. The result is weaker than
the one in Theorem 1, but no assumption concerning the computability of
time bounds has to be made.

When do we trust that a fast algorithm solves a given problem? At least for
well specified problems, like satisfiability, solving a
combinatoric puzzle, computing the digits of $\pi$, ..., we
usually invent algorithms, prove that they solve the problem and
in many cases also can prove good and fast time bounds. In these
cases, the provability assumptions in Theorem 1 are no real
restriction. The same holds for approximate algorithms which
guarantee a precision $\eps$ within a known time bound (many
numerical algorithms are of this kind). For exact/approximate
programs provably computing/converging to the right answer (e.g.\
traveling salesman problem, and also many numerical programs), but
for which no good, and easy to compute time bound exists, $M_{p*}$
is only optimal apart from a huge constant factor $5+d_p$ in time,
as discussed above. A precursor of algorithm $M_{p^*}$ for this
case, in a special setting, can be found in
\cite{Hutter:00f}\footnote{
The algorithm AI$\xi^{tl}$ creates an incremental policy for an
agent in an unknown non-Markovian environment, which is superior
to any other time $t$ and space $l$ bounded agent. The computation
time of AI$\xi^{tl}$ is of the order $t\!\cdot\!2^l$.}. For poorly
specified problems, Theorem 1 does not help at all.

\section{The Fast Algorithm $M_{p^*}$}\label{secFast}

One ingredient of algorithm $M_{p^*}$ is an enumeration of proofs of
increasing length in some formal axiomatic system. If a proof
actually proves that $p$ and $p^*$ are functionally equivalent and
$p$ has time bound $t_p$, add $(p,t_p)$ to a list $L$. The program
$p$ in $L$ with the currently smallest time bound $t_p(x)$ is
executed. By construction, the result $p(x)$ is identical to
$p^*(x)$. The trick to achieve the time bound stated in Theorem
1 is to schedule everything in a proper way, in order not to lose
too much performance by computing slow $p$'s and $t_p$'s before
{\it the} $p$ has been found.

To avoid confusion, we formally define $p$ and $t_p$ to be binary
strings. That is, $p$ is neither a program nor a function, but can
be informally interpreted as such. A formal definition of the
interpretations of $p$ is given below. We say ``p computes
function f'', when a universal reference Turing machine $U$ on
input $(p,x)$ computes $f(x)$ for all $x$. This is denoted by
$U(p,x)\!=\!f(x)$. To be able to talk about proofs, we need a
formal logic system
$(\forall,\lambda,y_i,c_i,f_i,R_i,\rightarrow,\wedge,=,...)$, and
axioms, and inference rules. A proof is a sequence of formulas,
where each formula is either an axiom or inferred from previous
formulas in the sequence by applying the inference rules.
See \cite{Fitting:96} 
or any other textbook on logic or proof theory. We only need to
know that {\it provability}, {\it Turing Machines}, and {\it
computation time} can be formalized:
\begin{enumerate}\parskip=0ex\parsep=0ex\itemsep=0ex
\item The set of (correct) proofs is enumerable.
\item A term $u$ can be defined such that the formula
$[\forall y\!:\!u(p,y)\!=\!u(p^*,y)]$ is true if, and
only if $U(p,x)\!=\!U(p^*,x)$ for all $x$, i.e. if
$p$ and $p^*$ describe the same function.
\item A term $tm$ can be defined such that the formula
$[tm(p,x)\!=\!n]$ is true if, and only if the computation time of
$U$ on $(p,x)$ is $n$, i.e.\ if $n\!=\!time_p(x)$.
\end{enumerate}
We say that $p$ is provably equivalent to $p^*$ if the formula
$[\forall y\!:\!u(p,y)\!=\!u(p^*,y)]$ can be proved. $M_{p^*}$
runs three algorithms $A$, $B$, and $C$ in parallel:

\pagebreak[0]
\begin{samepage}\begin{enumerate}\parskip=0ex\parsep=0ex\itemsep=0ex
\item[] {$\nq\nq$\bf Algorithm $M_{p^*}(x)$}
\item[] Initialize the shared variables $L:=\{\},\quad$
$t_{fast}:=\infty,\quad$ $p_{fast}:=p^*$.
\item[] Start algorithms $A$, $B$, and $C$ in parallel with
10\%, 10\% and 80\% \\
computational resources, respectively. \\
That is, $C$ performs 8 steps when $A$ and $B$ perform 1 step each.
\end{enumerate}\end{samepage}

\pagebreak[0]
\begin{samepage}\begin{enumerate}\parskip=0ex\parsep=0ex\itemsep=0ex
\item[] {$\nq\nq$\bf Algorithm $A$}
\item[] $\nq${\tt for} $i$:=1,2,3,... {\tt do}
\item[] pick the $i^{th}$ proof in the list of all proofs and\\
isolate the last formula in the proof.
\item[] {\tt if} this formula is equal to
$[\forall y\!:\!u(p,y)\!=u(p^*,y)\wedge u(t,y)\geq tm(p,y)]$ \\
for some strings $p$ and $t$, \\
{\tt then} add $(p,t)$ to $L$.
\item[] $\nq${\tt next} $i$
\end{enumerate}\end{samepage}

\pagebreak[0]
\begin{samepage}\begin{enumerate}\parskip=0ex\parsep=0ex\itemsep=0ex
\item[] {$\nq\nq$\bf Algorithm $B$}
\item[] $\nq${\tt for} all $(p,t)\!\in\!L$
\item[] run $U$ on all $(t,x)$ in parallel for all $t$ with relative
computational resources $2^{-\length(p)-\length(t)}$.
\item[] {\tt if} $U$ halts for some $t$ and $U(t,x)\!<\!t_{fast}$,\\
{\tt then} $t_{fast}:=U(t,x)$ and $p_{fast}:=p$.
\item[] $\nq${\tt continue $(p,t)$}
\end{enumerate}\end{samepage}

\pagebreak[0]
\begin{samepage}\begin{enumerate}\parskip=0ex\parsep=0ex\itemsep=0ex
\item[] {$\nq\nq$\bf Algorithm $C$}
\item[] $\nq${\tt for} k:=1,2,4,8,16,32,... {\tt do}
\item[] pick the currently fastest program $p:=p_{fast}$
        with time bound $t_{fast}$.
\item[] run $U$ on $(p,x)$ for $k$ steps.
\item[] {\tt if} $U$ halts in less than $k$ steps,
\item[] {\tt then} print result $U(p,x)$ and abort
        computation of $A$, $B$ and $C$.
\item[] $\nq${\tt continue $k$.}
\end{enumerate}\end{samepage}

\begin{figure}[tb]\epsfxsize=11cm
\centerline{\epsfbox{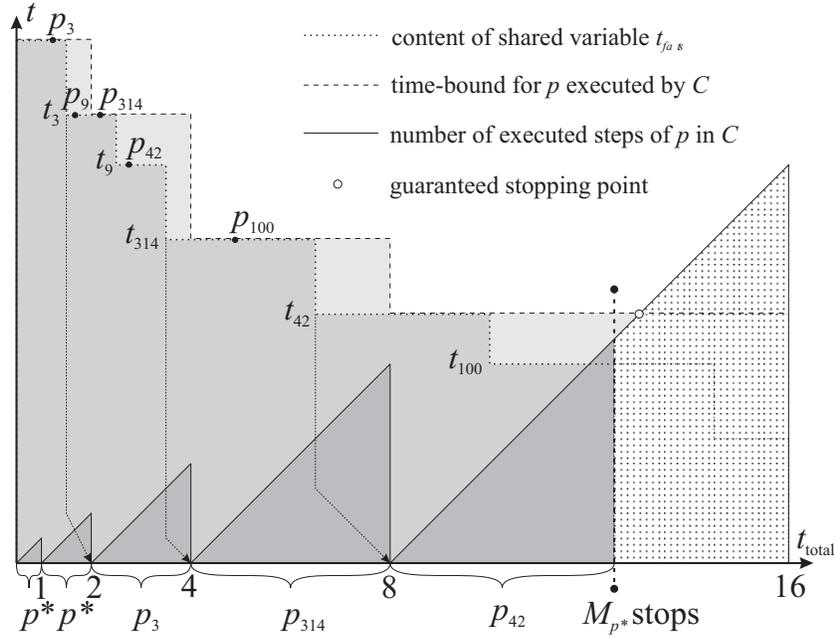}}
\caption{\label{figsch}Sample execution of $M_{p^*}$.
\it The horizontal axis marks the progress
of $M_{p^*}$. The vertical axis marks general time-like
quantities. The lower solid stepped curve is the content
of variable $t_{fast}$ at time $t$. Black dots on this curve
(denoted by $p$) mark events where algorithm $A$ creates and
adds another valid program $p$ together with its time-bound $t$ to
list $L$. The (later occurring) step in the curve marks the time
when algorithm $B$ has computed $t(x)$ and updates $t_{fast}$.
Algorithm $C$ starts $p$ at the next instruction step $k$ which is
a power of 2 and executes $p$ for $k$ steps. This is illustrated
by the exponentially increasing sawtooth curve. The time bound of
$p$ {\rm during execution} is indicated by the upper dashed stepped
curve. Algorithm $M_{p^*}$ terminates when the stepped curve crosses
the sawtooth curve (white circle), or earlier if the time-bound is not sharp.}
\end{figure}

Note that $A$ and $B$ only terminate when aborted by $C$. Figure
\ref{figsch} illustrates the time-scheduling on a fictitious
example. The discussion of the algorithm(s) in the following
sections clarifies details and proves Theorem 1.

\section{Time Analysis}\label{secTime}
Henceforth we return to the convenient abbreviations
$p(x)\!:=\!U(p,x)$ and $t_p(x)\!:=\!U(t_p,x)$. Let $p'$ be some
fixed algorithm that is provably equivalent to $p^*$, with
computation time $time_{p'}$ provably bounded by $t_{p'}$. Let
$\length(proof(p'))$ be the length of the binary coding of the,
for instance, shortest proof. {\it Computation time} always refers
to true overall computation time, whereas {\it computation steps}
refer to instruction steps. $steps=\alpha\!\cdot\!time$, if a
percentage $\alpha$ of computation time is assigned to an
algorithm.\\[-1.5ex]

\noindent A) To write down (not to invent!) a proof requires
$O(\length(proof))$ steps.
A time $O(N_{axiom}\!\cdot\!l(F_i))$ is needed to check whether a
formula $F_i$ in the proof $F_1F_2...F_n$ is an axiom, where
$N_{axiom}$ is the number of axioms or axiom-schemes, which is
finite. Variable substitution (binding) can be performed in linear
time. For a suitable set of axioms, the only necessary inference
rule is modus ponens. If $F_i$ is not an axiom, one searches for a
formula $F_j$, $j\!<\!i$ of the form $F_k\!\rightarrow\!F_i$ and
then for the formula $F_k$, $k\!<\!i$. This takes time
$O(l(proof))$. There are $n\leq O(l(proof))$ formulas $F_i$ to
check in this way. Whether the sequence of formulas constitutes a
valid proof can, hence, be checked in $O(\length(proof)^2)$ steps.
There are less than $2^{l+1}$ proofs of (binary) length $\leq\!l$.
Algorithm $A$ receives $\alpha\!=\!10\%$ of relative computation
time. Hence, for a proof of $(p',t_{p'})$ to occur, and for
$(p',t_{p'})$ to be added to $L$, at most time $T_A\leq{1\over
10\%}
\!\cdot\!2^{\length(proof(p'))+1}\!\cdot\!O(\length(proof(p'))^2)$
is needed. Note that the same program $p$ can and will be
accompanied by different time bounds $t_p$; for instance
$(p,time_p)$ will occur.\\[-1.5ex]

\noindent B) The time assignment of algorithm $B$ to the $t_p$'s
only works if the Kraft inequality $\sum_{(p,t_p)\in L}
2^{-\length(p)-\length(t_p)}\leq 1$ is satisfied \cite{Kraft:49}.
This can be ensured by using prefix free (e.g.\ Shannon-Fano)
codes \cite{Shannon:48,Li:97}. The number of steps to calculate
$t_{p'}(x)$ is, by definition, $time_{t_{p'}}(x)$. The relative
computation time $\alpha$ available for computing $t_{p'}(x)$ is
$10\%\!\cdot\!2^{-\length(p')-\length(t_{p'})}$. Hence,
$t_{p'}(x)$ is computed and $t_{fast}\!\leq\!t_{p'}(x)$ is checked
after time $T_B\leq T_A +
{1\over 10\%}\!\cdot\!2^{\length(p')+\length(t_{p'})} \!\cdot\!
time_{t_{p'}}(x)$. We have to add $T_A$, since $B$ has to wait, in
the worst case, time $T_A$ before it can start executing
$t_{p'}(x)$.\\[-1.5ex]

\noindent C) If algorithm $C$ halts, its construction guarantees that
the output is correct. In the following, we show that $C$ always
halts, and give a bound for the computation time.\\[-3.7ex]
\begin{enumerate}\parskip=0ex\parsep=0ex\itemsep=1ex
\item[\it i)] Assume that
algorithm $C$ stops before $B$ performed the check $t_{p'}(x)<t_{fast}$,
because a different $p$ already computed $p(x)$.
In this case $T_C\leq T_B$.
\item[\it ii)] Assume that $k=k_0$ in
$C$ when $B$ performs the check $t_{p'}(x)<t_{fast}$.
Running-time $T_B$ has passed until this point, hence
$k_0\leq 80\%\!\cdot\!T_B$ . Furthermore, assume that $C$ halts in
period $k_0$ because the program (different from $p'$) executed in
this period computes the result. In this case, $T_C\leq {1\over
80\%}2k_0\leq 2T_B$.
\item[\it iii)] If $C$ does not halt in period $k_0$
but $2k_0\!\geq t_{fast}$, then $p'(x)$ has enough time to compute
the solution in the next period $k=2k_0$, since $time_{p'}(x)\leq
t_{fast}\leq 4k_0-2k_0$. Hence $T_C\leq{1\over 80\%}4k_0\leq 4T_B$.
\item[\it iv)] Finally, if $2k_0\!<t_{fast}$ we ``wait''
for the period $k\!>\!k_0$ with $\odt k\!\leq\!t_{fast}\!<\!k$. In
this period $k$, either $p'(x)$, or an even faster algorithm,
which has in the meantime been constructed by A and B, will be
computed. In any case, the $2k-k>t_{fast}$ steps are sufficient to
compute the answer. We have ${80\%}\!\cdot\!T_C\leq 2k\leq
4t_{fast}\leq 4t_{p'}(x)$.
\end{enumerate}
The maximum of the cases {\it(i)} to {\it(iv)} bounds the
computation time of $C$ and, hence, of $M_{p^*}$ by
$$
  time_{M_{p^*}}(x) = T_C \;\leq\;
  \max\{4T_B,5t_{p}(x)\} \;\leq\;
  4T_B+5t_{p}(x) \;\leq\;
$$ $$
  \;\leq\; 5\!\cdot\!t_p(x) + d_p\!\cdot\!time_{t_p}(x) + c_p
$$ $$
  d_p=40\!\cdot\!2^{\length(p)+\length(t_p)},\quad
  c_p=40\!\cdot\!2^{\length(proof(p))+1}\!\cdot\!O(\length(proof(p)^2)
$$
where we have dropped the prime from $p$. We have also
suppressed the dependency of $c_p$ and $d_p$ on $p^*$ ($proof(p)$
depends on $p^*$ too), since we considered $p^*$ to be a fixed
given algorithm. The factor of 5 may be reduced to $4+\eps$ by
assigning a larger fraction of time to algorithm $C$. The
constants $c_p$ and $d_p$ will then be proportional to
$1\over\eps$. We were not able to further reduce this factor.

\section{Assumptions on the Machine Model}\label{secAssume}
In the time analysis above we have assumed that program simulation
with abort possibility and scheduling parallel algorithms can
be performed in real-time, i.e.\ without loss of performance.
Parallel computation can be avoided by sequentially performing all
operations for a limited time and then restarting all computations
in a next cycle with double the time and so on. This will increase
the computation time of $A$ and $B$ (but not of $C$!) by, at most,
a factor of $4$. Note that we use the same universal Turing
machine $U$ with the same underlying Turing machine model (number
of heads, symbols, ...) for measuring computation time of all
programs (strings) $p$, including $M_{p^*}$. This prevents us from
applying the linear speedup theorem (which is cheating
somewhat anyway), but allows the possibility of designing a $U$
which allows real-time simulation with abort possibility. Small
additive ``patching'' constants can be absorbed in the $O()$
notation of $c_p$. Theorem 1 should also hold for
Kolmogorov-Uspenskii
and Pointer machines.

\section{Algorithmic Complexity and the Shortest Algorithm}\label{secAIT}
Data compression is a very important issue in computer science.
Saving space or channel capacity are obvious applications. A less
obvious (but not far fetched) application is that of inductive inference
in various forms (hypothesis testing, forecasting, classification,
...). A free interpretation of Occam's razor is that the shortest
theory consistent with past data is the most likely to be correct.
This has been put into a rigorous scheme by \cite{Solomonoff:64}
and proved to be optimal in \cite{Solomonoff:78,Hutter:99}.
Kolmogorov Complexity is a universal notion of the information
content of a string \cite{Kolmogorov:65, Chaitin:66, Zvonkin:70}.
It is defined as the length of the shortest program computing
string $x$.
\beqn
  K_U(x) \;:=\; \min_p\{\length(p):U(p)=x\} \;=\; K(x) + O(1)
\eeqn
where $U$ is some universal Turing Machine. It can be shown
that $K_U(x)$ varies, at most, by an additive constant independent
of $x$ by varying the machine $U$. Hence, {\it the} Kolmogorov
Complexity $K(x)$ is universal in the sense that it is uniquely
defined up to an additive constant. $K(x)$ can be approximated from
above (is co-enumerable), but not finitely computable.
See \cite{Li:97} for an
excellent introduction to Kolmogorov Complexity and \cite{Li:00}
for a review of Kolmogorov inspired prediction schemes.

Recently, Schmidhuber \cite{Schmidhuber:01} has generalized
Kolmogorov complexity in various ways to the limits of
computability and beyond. In the following, we also need a
generalization, but of a different kind.
We need a short description of a function, rather than a string.
The following definition of the complexity of a function $f$
\beqn
  K'(f) := \min_p\{\length(p):U(p,x)=f(x)\,\forall x\}
\eeqn
seems natural, but suffers from not even being approximable.
There exists no algorithm converging to $K'(f)$, because it
is undecidable whether a program $p$ is the shortest program equivalent to a
function $f$.
Even if we have a program $p^*$ computing
$f$, $K'(p^*)$ is not approximable.
Using $K(p^*)$ is not a
suitable alternative, since $K(p^*)$ might be considerably longer
than $K'(p^*)$, as in the former case all information contained
in $p^*$ will be kept -- even that which is functionally
irrelevant (e.g.\ dead code). An alternative is to restrict ourselves to
provably equivalent programs. The length of the shortest one is
\beqn
  K''(p^*) \;:=\;
  \min_p\{\length(p): \mbox{a proof of }
  [\forall y\!:\! u(p,y)=u(p^*,y)] \mbox{ exists} \}
\eeqn
It can be approximated from above, since the set of all programs
provably equivalent to $p^*$ is enumerable.

Having obtained, after some time, a very short description $p'$ of
$p^*$ for some purpose (e.g.\ for defining a prior probability for
some inductive inference scheme), it is usually also necessary to
obtain values for some arguments. We are now concerned with the
computation time of $p'$. Could we get slower and slower
algorithms by compressing $p^*$ more and more? Interestingly this
is not the case. Inventing complex (long) programs is {\it not}
necessary to construct asymptotically fast algorithms, under the
stated provability assumptions, in contrast to Blum's Theorem
\cite{Blum:67,Blum:71}. The following theorem roughly says that there
is a {\em single} program, which is the fastest {\em and} the shortest
program.

\hfill

\begin{samepage}
{\sc Theorem 2.}
{\sl Let $p^*$ be a given algorithm or
formal specification of a function.
There exists a program $\tilde p$, equivalent
to $p^*$, for which the following holds
\beqn
\begin{array}{rl@{\;\leq\;}l}
  i)   & \length(\tilde p)
       & K''(p^*) + O(1) \\[1ex]
  ii)  & time_{\tilde p}(x)
       & 5\!\cdot\!t_p(x) + d_p\!\cdot\!time_{t_p}(x) + c_p
\end{array}
\eeqn
where $p$ is any program provably equivalent to $p^*$ with
computation time provably less than $t_p(x)$.
The constants $c_p$ and $d_p$ depend on $p$ but not on $x$.
} 
\end{samepage}

\hfill

To prove the theorem, we just insert the shortest algorithm $p'$
provably equivalent to $p^*$ into $M$, that is $\tilde p:=M_{p'}$.
As only $O(1)$ instructions are needed to build $M_{p'}$ from
$p'$, $M_{p'}$ has size $\length(p')\!+\!O(1)=K''(p^*)\!+\!O(1)$.
The computation time of $M_{p'}$ is the same as of $M_{p^*}$
apart from ``slightly'' different constants.

The following subtlety has been pointed out by Peter van Emde
Boas. Neither $M_{p^*}$, nor $\tilde p$ is {\em provably} equivalent
to $p^*$. The construction of $M_{p^*}$ in section \ref{secFast}
shows equivalence of $M_{p^*}$ (and of $\tilde p$) to $p^*$,
but it is a meta-proof which cannot be formalized within the
considered proof system. A formal proof of the correctness of
$M_{p^*}$ would prove the consistency of the proof system, which
is impossible by G{\"o}dels second incompleteness theorem. See
\cite{Hartmanis:79} for details in a related context.

\section{Generalizations}\label{secGeneral}

If $p^*$ has to be evaluated repeatedly, algorithm $A$ can be
modified to remember its current state and continue operation for
the next input ($A$ is independent of $x$!). The large offset time
$c_p$ is only needed on the first invocation.

$M_{p^*}$ can be modified to handle i/o streams, definable by a
Turing machine with monotone input and output tapes (and
bidirectional working tapes) receiving an input stream and
producing an output stream. The currently read prefix of the input
stream is $x$. $time_p(x)$ is the time used for reading $x$.
$M_{p^*}$ caches the input and output streams, so that algorithm
$C$ can repeatedly read/write the streams for each new $p$. The
true input/output tapes are used for requesting/producing a new
symbol . Algorithm $B$ is reset after $1,2,4,8,...$ steps (not
after reading the next symbol of $x$!) to appropriately take into
account increased prefixes $x$. Algorithm $A$ just continues. The
bound of Theorem 1 holds for this case too, with slightly increased
$d_p$.

The construction above also works if time is measured in terms of
the current output rather than the current input $x$. This measure
is, for example, used for the time-complexity
of calculating the $n^{th}$ digit of a computable real (e.g.\
$\pi$), where there is no input, but only an output stream.

\section{Summary \& Outlook}\label{secConc}
We presented an algorithm $M_{p^*}$ which accelerates the
computation of a program $p^*$. $M_{p^*}$ combines ($A$)
sequential search through proof space, ($B$) Levin search through
time-bound space, ($C$) and {\em sequential} program execution,
using a somewhat tricky scheduling. Under certain provability
constraints, $M_{p^*}$ is the asymptotically fastest algorithm for
computing $p^*$ apart from a factor 5 in computation time. Blum's
Theorem shows that the provability constraints are essential. We
have shown that the conditions on Theorem 1 are often, but not
always, satisfied for practical problems. For complex
approximation problems, for instance, where no good and fast time
bound exists, $M_{p^*}$ is still optimal, but in this case, only
apart from a large multiplicative factor. We briefly outlined how
$M_{p^*}$ can be modified to handle i/o streams and other
time-measures. An interesting and counter-intuitive consequence of
Theorem 1 was that the fastest program computing a certain
function is also among the shortest programs provably computing
this function. Looking for larger programs saves at most a finite
number of computation steps, but cannot improve the time order. To
quantify this statement, we extended the definition of Kolmogorov
complexity and defined two novel natural measures for the
complexity of a function. The large constants $c_p$ and $d_p$ seem
to spoil a direct implementation of $M_{p^*}$. On the other hand,
Levin search has been successfully applied to solve rather
difficult machine learning problems
\cite{Schmidhuber:97nn,Schmidhuber:97bias}, even though it suffers
from a large multiplicative factor of similar origin. The use of
more elaborate theorem-provers, rather than brute force
enumeration of all proofs, could lead to smaller constants and
bring $M_p^*$ closer to practical applications, possibly
restricted to subclasses of problems. A more fascinating (and more
speculative) way may be the utilization of so called transparent
or holographic proofs \cite{Babai:91}. Under certain circumstances
they allow an exponential speed up for checking proofs. This would
reduce the constants $c_p$ and $d_p$ to their logarithm, which is
a small value. I would like to conclude with a general question.
Will the ultimate search for asymptotically fastest programs
typically lead to fast or slow programs for arguments of practical
size? Levin search, matrix multiplication and the algorithm
$M_{p^*}$ seem to support the latter, but this might be due to our
inability to do better.

\subsection*{Acknowledgements}
Thanks to Monaldo Mastrolilli, J{\"u}rgen Schmidhuber, and Peter
van Emde Boas for enlightening discussions and for useful
comments. This work was supported by SNF grant 2000-61847.00.


\bibliographystyle{plain}

\end{document}